\RequirePackage{fix-cm}
\documentclass[smallextended]{svjour3}       
\smartqed  
\usepackage{graphicx}
\usepackage{amsmath}
\usepackage{amssymb}
\usepackage{dsfont}
\usepackage{xcolor}

\usepackage[misc,geometry]{ifsym}
\usepackage{etoolbox}
\newcommand*{\affaddr}[1]{#1} 
\newcommand*{\affmark}[1][*]{\textsuperscript{#1}}

\newcommand{\ket}[1]{\ensuremath\,|{#1}\rangle}
\newcommand{\bra}[1]{\ensuremath\langle #1 |\,}

\newcommand{\id}{\mathds 1}
\begin{document}


\title{Distributing Entanglement with Separable States: Assessment of Encoding and Decoding Imperfections}

\titlerunning{Distributing Entanglement with Separable States}        

\author{Hannah McAleese\protect\affmark[1]	    \and
        Gediminas Juska\affmark[2]			\and
        Iman Ranjbar Jahromi\affmark[2]	\and
        Emanuele Pelucchi\affmark[2]		\and
        Alessandro Ferraro\affmark[1]		\and
        Mauro Paternostro\affmark[1] 
}

\institute{\Letter \quad Hannah McAleese\\
\email{hmcaleese02@qub.ac.uk}\\
\\
\affaddr{\affmark[1]School of Mathematics and Physics, Queen's University, Belfast BT7 1NN, United Kingdom}\\
\\
\affaddr{\affmark[2]Epitaxy and  Physics  of Nanostructures, Tyndall  National  Institute, University College  Cork, Lee Maltings, Dyke Parade, Cork, Ireland}
}

\authorrunning{Hannah McAleese et al.} 


\date{Received: 22 January 2021 / Accepted: 18 May 2021}

\maketitle

\begin{abstract}
Entanglement can be distributed using a carrier which is always separable from the rest of the systems involved. Up to now, this effect has predominantly been analyzed in the case where the carrier-system interactions take the form of ideal unitary operations, thus leaving untested its robustness against either non-unitary or unitary errors. We address this issue by considering the effect of incoherent dynamics acting alongside imperfect unitary interactions. In particular, we determine the restrictions that need to be placed on the interaction time, as well as the strength of the incoherent dynamics. We find that with non-unitary errors, we can still successfully distribute entanglement, provided we measure the carrier in a suitable basis. Introducing imperfections in the unitary dynamics, we show that entanglement gain is possible even with substantial unitary errors. Moreover, certain variations in the strength of the unitary dynamics can allow for greater robustness against non-unitary errors. Therefore, even in experimental settings where unitary operations cannot be carried out without imperfections, it is still possible to generate entanglement between two systems using a separable carrier.
\end{abstract}

\section{\label{sec:intro}Introduction}

Quantum entanglement is one of the features that makes quantum mechanics so fascinating and counter-intuitive. Furthermore, not only is quantum entanglement of fundamental interest but it is also a valuable resource in quantum information, as it is instrumental for applications such as teleportation \cite{teleportation}, dense coding \cite{denseCoding}, and quantum key distribution \cite{qkd,acin07}. The challenge that has to be faced lies in exploiting this resource despite its fragility. In particular, in quantum communication or in computation, we would like to generate and sustain entanglement as and when it is needed. One way of achieving this is through entanglement distribution. 

\begin{figure}[t]
	\centering
	\includegraphics[width=\columnwidth]{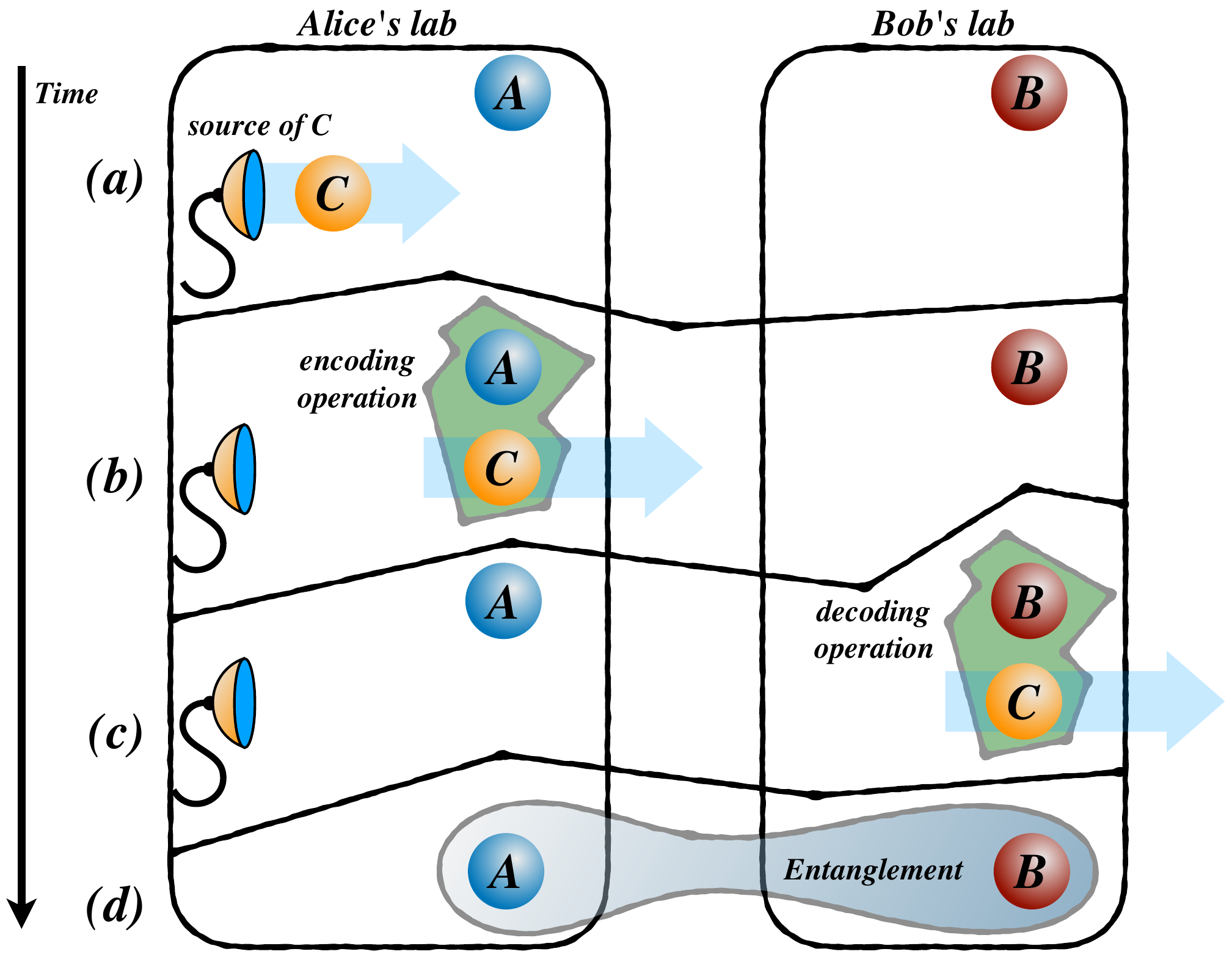}
	\caption{{\bf (a)} Carrier $C$ is emitted from a source in Alice's lab. {\bf (b)} Encoding operation: Alice's qubit and the carrier interact in Alice's lab before she sends the carrier to Bob. This interaction generates entanglement between $A$ and $BC$. {\bf (c)} Decoding operation: Bob's qubit and the carrier interact, generating entanglement between $B$ and $AC$. {\bf (d)} The result of the two operations: entanglement is generated between Alice and Bob's qubits $A$ and $B$.}
	\label{fig:EDSSdiagram}
\end{figure}

Entanglement can be distributed either directly or indirectly. Direct entanglement distribution between two parties (conventionally, Alice and Bob) involves Alice creating an entangled state of two systems in her laboratory and sending one to Bob. Indirect entanglement distribution, which is the focus of this paper, generates entanglement between a system $A$ in Alice's laboratory and another in Bob's laboratory $B$ through sending a carrier system $C$ as in Fig. \ref{fig:EDSSdiagram}. The carrier first interacts with Alice's system generating entanglement in the bipartition $A|BC$. This interaction is called the encoding operation. Then the carrier is sent to Bob and interacts with his system, localising the entanglement onto Alice and Bob's systems only. We call this interaction the decoding operation.

Surprisingly, entanglement can be distributed indirectly using a carrier that remains separable from the two systems throughout the process. A scheme through which such a task can be achieved was first proposed by Cubitt {\it et al.} in Ref.~\cite{cubitt} for the case of discrete variables and later extended to continuous variables in Refs.~\cite{mista08,mista09,mista13}. Prototypes of such schemes have been demonstrated experimentally for both discrete and continuous variables \cite{fedrizzi13,vollmer13,peuntinger13}. So far, several aspects of entanglement distribution via separable states (EDSS) have been studied. It was found that, though entanglement is not needed between the carrier and the other systems, quantum discord has a key role to play in EDSS~\cite{alphaPaper,streltsov12}. This is fundamentally relevant --- in light of the current quest to clarify the potential resource-like role of discord in quantum information processing \cite{bera17} --- and practically interesting, as discord appears to be much more robust than entanglement to environmental effects~\cite{werlang09,ferraro10,mazzola10,wang10,fanchini10}. The initial states of $AB$ that can be used successfully for EDSS have been found~\cite{kay12}, and the ways in which different entanglement measures, different noisy channels and amounts of initial correlations affect entanglement distribution have been analysed~\cite{streltsov15}. The concept of excessive entanglement distribution was introduced in Ref.~\cite{zuppardo16} as a protocol where the entanglement gained between $A$ and $B$ is greater than the entanglement between $C$ and $AB$. Needless to say, any EDSS protocol is excessive. Multipartite generalization of EDSS have been put forward~\cite{karimipour15} and the effect of noise on this process was studied~\cite{bordbar18}. More recently, it has been realized that EDSS can be used to detect non-classicality in inaccessible objects by testing if they can be used as the carrier in this process~\cite{krisnanda17}, a result that can be used, in principle, to infer the potential quantum nature of certain biological processes~\cite{krisnanda2018} and gravity~\cite{krisnanda2019}.

Up to now, studies have predominantly been carried out in the case where the interactions in the encoding and decoding steps take the form of a unitary operation. However, the impact of imperfections in the encoding and decoding steps on the performance of the protocol is yet unexplored. The general context in the case of imperfect encoding and decoding would see $C$ interacting with $A$ and then $B$ through incoherent mechanisms. This is a relevant point to address at both the fundamental and experimental level. On one hand, the experimental implementation of encoding and decoding operations, which are central to the performance of EDSS, is unlikely to be exempt from imperfections that make the assumption of unitarity untenable. This is the case, for instance, of promising setups for the test of EDSS, namely cavity optomechanics in a membrane-in-the-middle configuration~\cite{AKM}, quantum spin chains~\cite{Baart17,Sahling15}, and semiconductor-based quantum photonics~\cite{chung16}, where only an open-system map would appropriately describe the dynamics. Recently, entanglement distribution to non-interacting optical fields mediated by a mechanical mode has been reported in a setting that is suggestive of EDSS performance~\cite{barzanjeh19}. On the other hand, there is no analysis of the performance of EDSS under imperfect encoding and decoding operations, and it is worth exploring the robustness of such a scheme under only partially coherent operations.

The remainder of this paper is organized as follows. In Sec.~\ref{sec:setup} we describe the system under scrutiny and the tools that we used to quantify entanglement. In Sec.~\ref{sec:CABcond} we explain the conditions needed to ensure that the carrier remains separable from the rest of the system throughout the process. In Sec.~\ref{sec:doesntPreventEDSS} we show that we can indeed still distribute entanglement with separable states and incoherent dynamics. Sec.~\ref{sec:varyUnitaryStrength} is devoted to the analysis of the effects induced by the changes in the strength of the coherent dynamics. Finally in Sec.~\ref{sec:conc} we summarise our findings and present our conclusions. 

\section{\label{sec:setup}Entanglement distribution protocol}

In what follows, we denote $E_{A_1|A_2}$ the entanglement of a state $\rho$ with respect to the bipartition $A_1|A_2$. Such entanglement will be quantified using the negativity \cite{negativity}, 
\begin{equation} 
	E_{A_1|A_2} = \frac{||\rho^{T_{A_1}}||-1}{2},
\end{equation}
where $\rho^{T_{A_1}}$ is the partial transposition of $\rho$ with respect to $A_1$, while $||\rho|| = \mathrm{Tr} \sqrt{\rho^\dagger \rho}$ is the trace norm. {A maximally entangled state of qubits $A_1$ and $A_2$ would have negativity $E_{A_1|A_2}=1/2$.}

\subsection{Original Protocol Proposed by Cubitt {\it et al.}}

Our entanglement distribution protocol builds on the example of Cubitt {\it et al.} in Ref.~\cite{cubitt}. They use CNOT gates to entangle Alice's system $A$ and Bob's system $B$ via a carrier qubit $C$. Starting with an appropriate separable initial state, they were able to generate $AB$ entanglement while the carrier qubit $C$ shared no entanglement with $A$ or $B$ throughout the process. The initial state, which we will label $\Lambda_\mathrm{sep}$, is
\begin{equation} \label{eq:cubittInitial}
\begin{aligned}
    \Lambda_\mathrm{sep} &= \frac{1}{6} \sum_{k=0}^3 \ket{\Psi_k}\bra{\Psi_k}_A \otimes \ket{\Psi_{-k}}\bra{\Psi_{-k}}_B \otimes \ket{0}\bra{0}_C \\ &+ \frac{1}{6} \sum_{i=0}^1 \ket{i}\bra{i}_A \otimes \ket{i}\bra{i}_B \otimes \ket{1}\bra{1}_C,
\end{aligned}
\end{equation}
where $\ket{\Psi_k} = (\ket{0} + e^{ik\pi/2} \ket{1})/\sqrt{2}$. Written in this way, we can see that $\Lambda_\mathrm{sep}$ fits the definition of a mixed tripartite separable state.

The first step of the protocol, or the encoding operation, involves acting on $\Lambda_\mathrm{sep}$ with a CNOT operation on $AC$ where $A$ is the control qubit. This results in the state
\begin{equation} \label{eq:cubittGHZ}
    \sigma = \frac{1}{3} \ket{\mathrm{GHZ}}\bra{\mathrm{GHZ}}_{ABC} + \frac{1}{6} \id_A \otimes \big( \ket{01}\bra{01} + \ket{10}\bra{10} \big)_{BC}
\end{equation}
where $\ket{\mathrm{GHZ}} = (\ket{000} + \ket{111})/\sqrt{2}$ is a Greenberger-Horne-Zeilinger state. Now the state has become entangled in the bipartition $A|BC$ with negativity $E_{A|BC} = 1/6$. However, as the CNOT acts only on $A$ and $C$, there is no entanglement generated yet between $B$ and $AC$. Importantly, the carrier $C$ remains separable from $AB$ since the state $\sigma$ is invariant under permutations of $B$ and $C$.

The next step is another CNOT operation but this time acting on $B$ and $C$ with $B$ the control qubit. We call this step the decoding operation. The final state is then
\begin{equation} \label{eq:cubittFinal}
    \tau = \frac{1}{3} \ket{\phi^+}\bra{\phi^+}_{AB} \otimes \ket{0}\bra{0}_C + \frac{2}{3} \id_{AB} \otimes \ket{1}\bra{1}_C,
\end{equation}
where $\ket{\phi^+}=(\ket{00}+\ket{11})/\sqrt{2}$ is a maximally entangled Bell state. The two systems $A$ and $B$ are now entangled as desired. It is easy to see that the carrier $C$ is still separable from $AB$. Thus Cubitt {\it et al.} achieved entanglement distribution with separable states.

\subsection{Modified Protocol}

In our work, we make some changes to Cubitt's protocol. The first is to relax the condition that the initial state is separable in any partition. Instead, we follow Example 2 in Ref.~\cite{alphaPaper}, where we may choose to have some initial entanglement between $A$ and $B$. The initial state is $\alpha_{ABC} (p) = p \Lambda_\mathrm{sep} + (1-p) \Lambda_\mathrm{ent}$, where 
\begin{equation}
	\label{cubitsep}
	\begin{aligned}
		\Lambda_\mathrm{sep} &=\frac{1}{6}\left[ \left(2  \ket{\phi^+} \bra{\phi^+} {+} {\ket{01} \bra{01} {+} \ket{10} \bra{10}}  \right)_{AB} \otimes \ket{0} \bra{0}_{C}\right. \\ 
		&\left.+ \left( \ket{00} \bra{00} {+}\ket{11} \bra{11} \right)_{AB} \otimes \ket{1} \bra{1}_{C}\right],\\
		\Lambda_\mathrm{ent} &= \frac{1}{3}\left[ \left(  \ket{00} \bra{00} +  \ket{11} \bra{11} \right)_{AB} \otimes \ket{1} \bra{1}_{C} \right. \\ 
		&\left.+\ket{\phi^+}\bra{\phi^+}_{AB}  \otimes \ket{0} \bra{0}_{C}\right],
	\end{aligned}
\end{equation}
where we have rewritten the state $\Lambda_\mathrm{sep}$ from Eq.~(\ref{eq:cubittInitial}) in a simplified manner. The state $\Lambda_\mathrm{ent}$ is separable with respect to the $C|AB$ bipartition but entangled in any other one with negativity $E_{A|BC} = E_{B|AC} = 1/6$. State $\alpha_{ABC}(p)$ depends on parameter $p \in [0,1]$, which determines the amount of initial entanglement between $A$ and $B$. As $p$ increases, entanglement decreases.


It is useful to emphasise that our choice of initial state affects the results of our analysis. There are other initial states of $ABC$ which would allow for EDSS and they could be investigated in an equally righteous manner. However, as we are interested in adapting the example and protocol given by Cubitt {\it et al.} in Ref.~\cite{cubitt}, which kickstarted the investigation on EDSS, we have restricted our study to this particular state. On a related note, it is worth remarking that there is no strict necessity to  choose CNOT gates for the encoding and decoding operations to successfully achieve EDSS. It is a sensible choice firstly due to the entangling power of CNOT operations~\cite{Zanardi} and secondly as they are widely used, especially in quantum circuits and algorithms. However, though we follow the protocol in Ref.~\cite{cubitt}, other approaches have included using continuous-variable states with beam splitters as the encoding and decoding operations~\cite{mista08,mista09} or allowing $A$, $B$ and $C$ to interact continuously and describing the dynamics with a single interaction Hamiltonian~\cite{cubitt,krisnanda17}.

The next change we make to the protocol in Ref.~\cite{cubitt} is to add incoherence to the encoding operation. A CNOT operation is generated by the Hamiltonian $H_{AC}=\lambda \vert 1\rangle\langle 1\vert_A\otimes(\sigma^x_C-\id_C)$ so that the unitary operation $U_{AC} = e^{-i H_{AC} t}$ is equal to the desired CNOT operation when the interaction time $t=\pi/(2\lambda)$. The frequency $1/t$ sets a scale for any other rate or interaction strength involved in our analysis. We add incoherence to the dynamics by introducing an excitation-exchange term of the form $\sigma_A^+\sigma_C^-+h.c.$ with $\sigma^+_j = \ket{1} \bra{0}_j$ and $\sigma^-_j = \ket{0} \bra{1}_j$ the ladder operators for qubit $j=A,C$. We thus modify the encoding step from the unitary CNOT transformation to the map described by the master equation  
\begin{equation} 
	\label{eq:encoding}
	\dot{\rho} = - i [H_{AC}, \rho] + \gamma_{AC} \mathcal{L}_{AC} (\rho)
\end{equation}
where $\mathcal{L}_{AC} (\rho) = 2 O_{AC}\rho O^\dag_{AC}  - \{O^\dag_{AC}O_{AC},\rho\}$ is the Lindblad superoperator that describes the incoherent energy exchange between $A$ and $C$ with $O_{AC}=\sigma^+_{A} \sigma^-_{C}$. The strength of the incoherent dynamics is $\gamma_{AC}$ and we allow this interaction to take place for a time $t_{AC}$, so that $t$ varies from 0 to $t_{AC}$. The encoding step should result in increased entanglement between $A$ and $BC$ while keeping $C$ and $AB$ separable, as in Ref.~\cite{cubitt}.

We make the same changes to the decoding operation between $B$ and $C$. We assume this to occur in a similar way to the encoding operation, i.e. according to the dynamical map
\begin{equation} \label{eq:decoding}
	\dot{\rho} = - i [H_{BC}, \rho] + \gamma_{BC} \mathcal{L}_{BC} (\rho),
\end{equation}
where $H_{BC}$ is the Hamiltonian generating the ${\rm CNOT}_{BC}$ operation that decodes information carried by $C$ and ${\cal L}_{BC}$ is analogous to the Lindblad superoperator invoked for the encoding step. 
We let $B$ and $C$ evolve according to Eq.~\eqref{eq:decoding} for a time $t_{BC}$. During the decoding step, $B$ and $AC$ should become entangled \cite{cubitt,alphaPaper}. 

Upon completion of such operations, we aim to find the amount of entanglement gained by $A$ and $B$ from such distribution process. We will evaluate the negativity between $A$ and $B$ both when the carrier $C$ is traced out and when it is acted on by a projective measurement. Our aim is to find positive entanglement gain between $A$ and $B$ while the bipartition $C|AB$ remains separable for the duration of the protocol.

A remark is due: the adoption of a Lindblad-like description of the incoherent part of the dynamics calls for the validity of Born-Markov assumptions on the mechanism responsible for this part of the dynamics of the system. This obviously sets constraints on the physical settings to be used in order to provide an embodiment of the EDSS protocol itself. While we refer to Sec.~\ref{sec:conc} for a brief mention of a plausible and actually very promising experimental platform, here we want to stress that the scopes of our study are to provide a {\it universal} analysis, i.e. a study providing predictions of broad applicability and unrelated to the specific details of any chosen experimental configuration. This implies that, provided that the chosen settings satisfy the conditions for the validity of a Lindblad-like master equation, our study will give valuable information on the performance of the EDSS protocol.

\section{Conditions imposed by the carrier separability requirement} \label{sec:CABcond}

In this Section we determine the conditions that should be satisfied in order for $C$ to remain separable from $AB$. We will first study the effect of interaction times $t_{AC}$ and $t_{BC}$ before analysing the impact that the strengths of the incoherent dynamics $\gamma_{AC}$ and $\gamma_{BC}$ have on the amount of entanglement $E_{C|AB}$ set upon completing the distribution protocol.

\subsection{Conditions on the interaction times}\label{sec:times}

Firstly, looking at the case where both $t_{AC}$ and  $t_{BC}$ grow indefinitely, we find that for both the encoding and decoding steps, there is no unique steady state. For instance, states $\rho = \ket{000} \bra{000}$ and $\rho = \ket{001} \bra{001}$ are both steady states of the encoding step in Eq. (\ref{eq:encoding}) and states $\rho = \ket{000} \bra{000}$ and $\rho = \ket{100} \bra{100}$ are steady states of the decoding operation in Eq.~(\ref{eq:decoding}). Therefore, we must select a specific initial state and investigate how the state changes as the time evolves.

We now turn our attention to states produced when an initial state of the form $\alpha_{ABC} (p)$ evolves according to Eq.~(\ref{eq:encoding}) for a large time $t_{AC}$. The unitary dynamics is determined by the Hamiltonian defined in Sec.~\ref{sec:setup} and is therefore given by
\begin{equation}
		U_{AC} = \ket{0} \bra{0}_A \otimes \id_C + 
		 \ket{1} \bra{1}_A \otimes \frac{1}{2} \left[ (1+e^{2 i \lambda t}) \id_C + (1-e^{2 i \lambda t} ) \sigma^x_C \right].
\end{equation}
As the interaction time grows, the operation being realized alternates between a CNOT and an identity operator. Therefore, the state of the system at large time varies periodically, with period $\pi/\lambda$. As the CNOT operation has the effect of entangling the system with respect to bipartition $A|BC$, it is beneficial for our purposes to allow the system to evolve for as long as is needed for a full CNOT operation to be carried out. Setting $\lambda = \pi/2$, we thus take the interaction time $t_{AC}$ to be finite and belonging to the set $t_{AC} \in \{1, 3, 5, \dots$\} so that the evolution of state $\alpha_{ABC} (p)$ with respect to the unitary dynamics is equivalent to acting on the state with a CNOT operation. Similarly, we restrict $t_{BC}$ to the set $t_{BC}\in \{1, 3, 5, \dots$\} to implement a full CNOT operation in the decoding operation.

\subsection{Conditions on incoherent interaction strengths with increasing interaction times}

\begin{figure}[t]
	\centering
	{\bf (a)}\hskip6cm{\bf (b)}
	\includegraphics[width=\columnwidth]{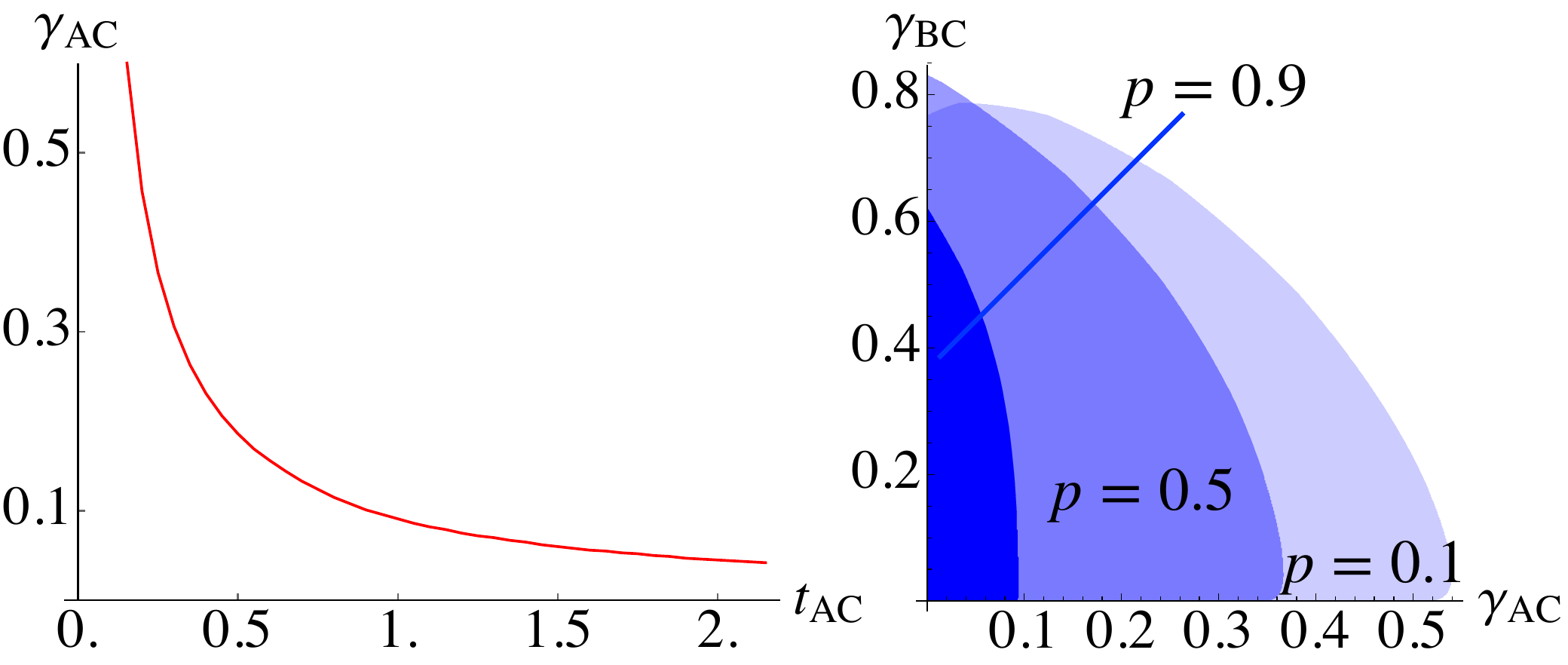}
	\caption{Panel {\bf (a)} shows the maximum value that $\gamma_{AC}$ can take so that $E_{C|AB}=0$ after the encoding operation as the interaction time $t_{AC}$ increases. The initial state is taken to be $\alpha_{ABC}(p)$ with $p=0.9$ and the parameter in the Hamiltonian is fixed at $\lambda = \pi/2$ so that the CNOT operation is realized at time $t_{AC}=1$. Panel {\bf (b)} reports the strengths of incoherent dynamics for which $E_{C|AB}=0$ throughout the whole entanglement distribution process. In these simulations we have taken assumed the initial state $\alpha_{ABC}(p)$ with $p=0.9$ (dark blue), $p=0.5$ (blue) and $p=0.1$ (light blue).}
	\label{fig:steadyState}
\end{figure}

It remains to investigate, then, just how long we can allow $A$ and $C$ to interact before the carrier system $C$ becomes entangled with $AB$. To do this, we study the restrictions that increasing interaction time imposes on the incoherent interaction strength. Fig.~\ref{fig:steadyState} {\bf (a)} shows the maximum value that $\gamma_{AC}$ can take in order for the state of the system to remain separable with respect to the $AB\vert C$ bipartition during the encoding operation. We find that we must compromise between the length of time of the interaction between $A$ and $C$ and the strength of the incoherent part of the dynamics. Our initial state is designed so as to allow $A$ and $BC$ to become entangled through a CNOT operation without the bipartition $C|AB$ being affected. We achieve this by adding the projectors $\ket{001}\bra{001}$ and $\ket{111}\bra{111}$ to the initial state as can be seen in Eq.~(\ref{cubitsep}). However, the superoperator $\mathcal{L}_{AC}$ involves an excitation exchange mechanism which diminishes the contribution of projector $\ket{001}\bra{001}$ in the state as the system evolves. This leads to unwanted entanglement between $C$ and $AB$, which increases as $t_{AC}$ and $\gamma_{AC}$ increase.

Therefore, a long interaction time $t_{AC} > 1$ restricts the range of values of $\gamma_{AC}$ that we can study if we are to keep $C$ and $AB$ separable. We must choose between having the possibility to choose from an ample range of values of the interaction time or incoherent strengths. As our aim is to analyse the effect of incoherent dynamics on the entanglement produced through the protocol, we choose to restrict the interaction time. As a result, we fix the value of such parameter to be the least possible allowing a CNOT operation to be performed, so that we can assess a broader range of values of $\gamma_{AC}$. Therefore, in the remainder of our analysis we fix the interaction times to be $t_{AC}=t_{BC}=1$. 

\subsection{\label{sec:encodingMoreRestricted} Conditions on the incoherent interaction strengths when interaction times are fixed}

Now that we have set an interaction time for the encoding and decoding dynamics, we can study the restrictions we must place on the strengths of the incoherent dynamics $\gamma_{AC}$ and $\gamma_{BC}$. Though the processes in Eqs.~(\ref{eq:encoding}) and (\ref{eq:decoding}) mirror one another, the values that $\gamma_{AC}$ can take to preserve the effectiveness of the protocol are much more constrained than those for $\gamma_{BC}$. This is clearly visible from Fig.~\ref{fig:steadyState} {\bf (b)}, which shows the region of values of $\gamma_{AC}$ and $\gamma_{BC}$ that allow for a separable carrier for different initial states.

This asymmetry is due to the differences in the state before the encoding operation and after the decoding one. Initially the state has a contribution of the form $\ket{\phi^+} \bra{\phi^+}_{AB} \otimes \ket{0} \bra{0}_{C}$ as in Eq.~(\ref{cubitsep}). The CNOT operation acting on $A$ and $C$ in the encoding step changes this to a GHZ state $\ket{\mathrm{GHZ}} \bra{\mathrm{GHZ}}_{ABC}$ as we can see in Eq.~(\ref{eq:cubittGHZ}). The additional projectors in Eq.~(\ref{cubitsep}) ensure that the state is separable with respect to the bipartition $AB|C$ after the encoding operation. As GHZ states are genuinely tripartite entangled, a change in the added projectors could easily entangle the state. However, the CNOT operation on $B$ and $C$ in the decoding step has the effect of reversing the previous transformation; this part of the state returns to the form of a tensor product between a Bell state of $AB$ and $\ket{0} \bra{0}_{C}$. Clearly $C$ is separable from $AB$ after this operation. As a result, the changes in projectors during the decoding operation have a weaker effect on the entanglement of $C$ and $AB$. This enables us to be much more flexible in our choice of incoherent strength in the decoding operation in comparison with the encoding one. Despite this, the excitation exchange between $B$ and $C$ still has the effect of coupling $C$ with $AB$ when the unitary dynamics is not equivalent to a full CNOT operation. For this reason, we are still limited in the range of suitable values for $\gamma_{BC}$; a sufficiently strong incoherent dynamics will produce entanglement in the bipartition $C|AB$ at some point during the decoding step of the protocol.

We also find that the lower the initial amount of entanglement, the smaller the area of the region where successful EDSS is achieved. For a separable initial state (i.e. for $p=1$), we can only achieve EDSS when $\gamma_{AC}=0$, i.e. the encoding operation is entirely unitary, and {$\gamma_{BC}\in[0,0.006]$. This effect is again due to the contribution of the projector $\ket{001} \bra{001}$ which is required to keep the carrier $C$ separable from $AB$. Even a slight decrease in the value taken by the probability to have element $\ket{001} \bra{001}$ in state $\Lambda_\mathrm{sep}$ [cf. Eq.~(\ref{cubitsep})] results in entanglement between $C$ and $AB$. However, the contribution of the same projector to $\Lambda_\mathrm{ent}$  is twice as large as in $\Lambda_\mathrm{sep}$. Therefore, the lower the value of $p$, the higher the contribution from $\ket{001} \bra{001}$ and the more robust the state is against entanglement forming between $C$ and $AB$.} 

In order to avoid unnecessary restrictions, from now on we will focus on $p=0.9$, i.e. a situation where a relatively small initial entanglement is present in the state ($E_{A|BC}=0.0167$). We are therefore considering values of $\gamma_{AC}$ between 0 and 0.0945 and $\gamma_{BC}$ between 0 and 0.622 as displayed in Fig.~\ref{fig:steadyState} {\bf (b)}.

\section{Effect of the Incoherent Dynamics On Entanglement} \label{sec:doesntPreventEDSS}

Now that we have found the conditions under which the encoding and decoding interactions leave the bipartition $C$-vs-$AB$ separable, we must determine whether or not entanglement can still be distributed. 

\begin{figure}[t]
	\centering
	\includegraphics[width=0.8\columnwidth]{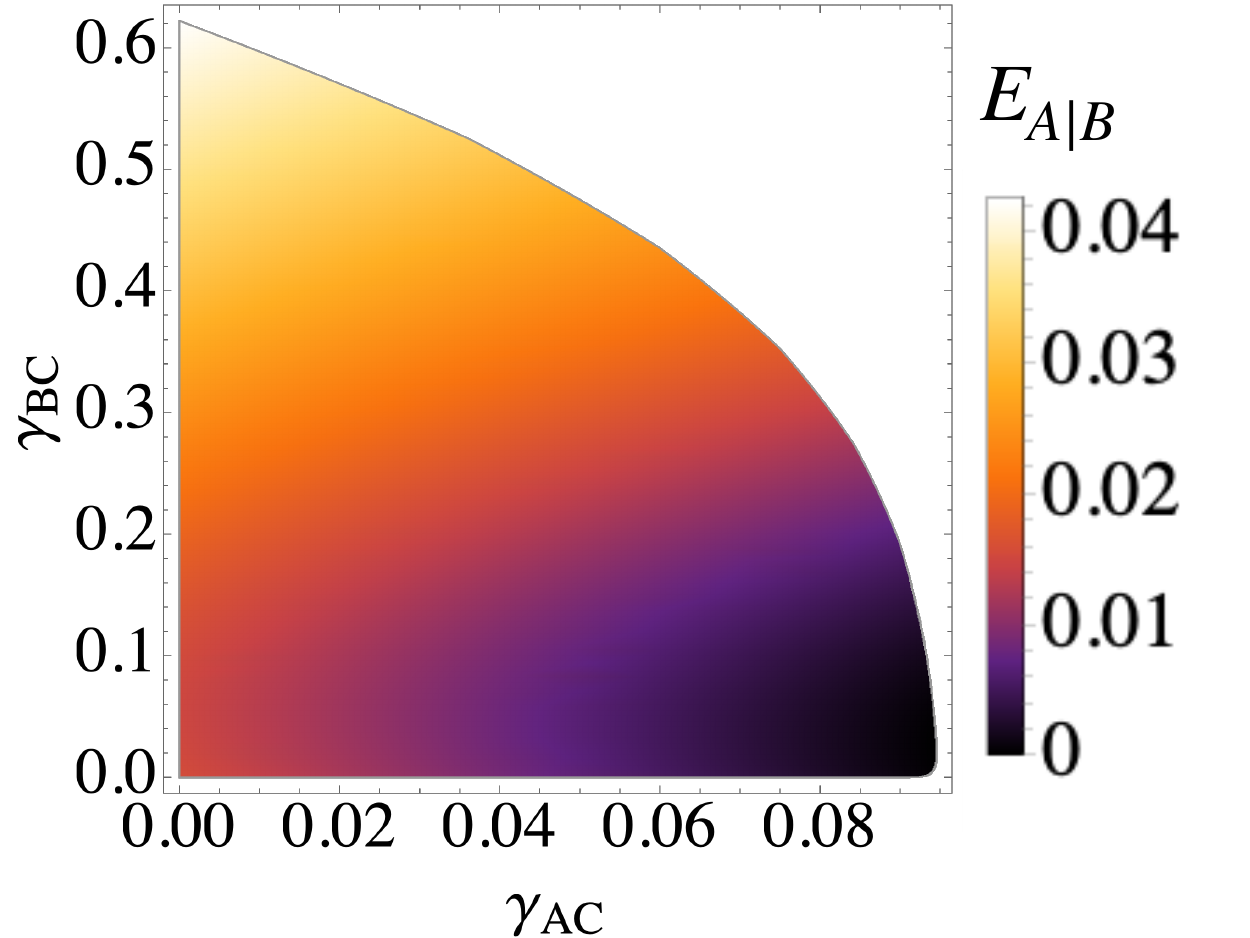}\\
	\caption{Entanglement between $A$ and $B$ if $C$ is traced out of the system. Although the amount of entanglement is very small, the incoherent dynamics in the decoding step are advantageous in this setting. Since the initial state has $E_{A|B}=0.0167$, it is possible to lose entanglement during the protocol using this method.}
	\label{fig:measTraceC}
\end{figure}

In Sec.~\ref{sec:encodingMoreRestricted}, we have established that it is possible to carry out EDSS for a certain range of values of $\gamma_{AC}$ and $\gamma_{BC}$. However, we do not yet know how much entanglement we generate between $A$ and $B$ during such a process. Fig.~\ref{fig:measTraceC} shows the entanglement between $A$ and $B$ at the end of the protocol when the carrier is simply traced out of the system. Interestingly, the incoherent dynamics in the decoding step of the process prove to be beneficial; the stronger the incoherent dynamics in the decoding step $\gamma_{BC}$, the higher the entanglement generated. To illustrate why such increase occurs, it is sufficient to consider the final state $\tau$ of the protocol with unitary dynamics in Eq.~(\ref{eq:cubittFinal}). One contribution comes from an entangled Bell state, which is unaffected by the action of the Lindblad superoperator $\mathcal{L}_{BC}$. The other comes from an identity operator, which is required for the state to be separable after $C$ is traced out of the system. The action of $\mathcal{L}_{BC}$ on the state causes the population of $\ket{10}$ in the identity to be transferred to $\ket{11}$. Consequently, as $\gamma_{BC}$ increases, the contribution from the identity in $\tau$ decreases and the entanglement between $A$ and $B$ grows. Nevertheless, the amount of entanglement produced in this way is clearly very small. In addition, note that the initial entanglement between $A$ and $B$ is 0.0167. The entanglement gained since the beginning of the protocol is therefore very small and can even be negative.

\begin{figure*}[t]
	\includegraphics[width=.95\linewidth]{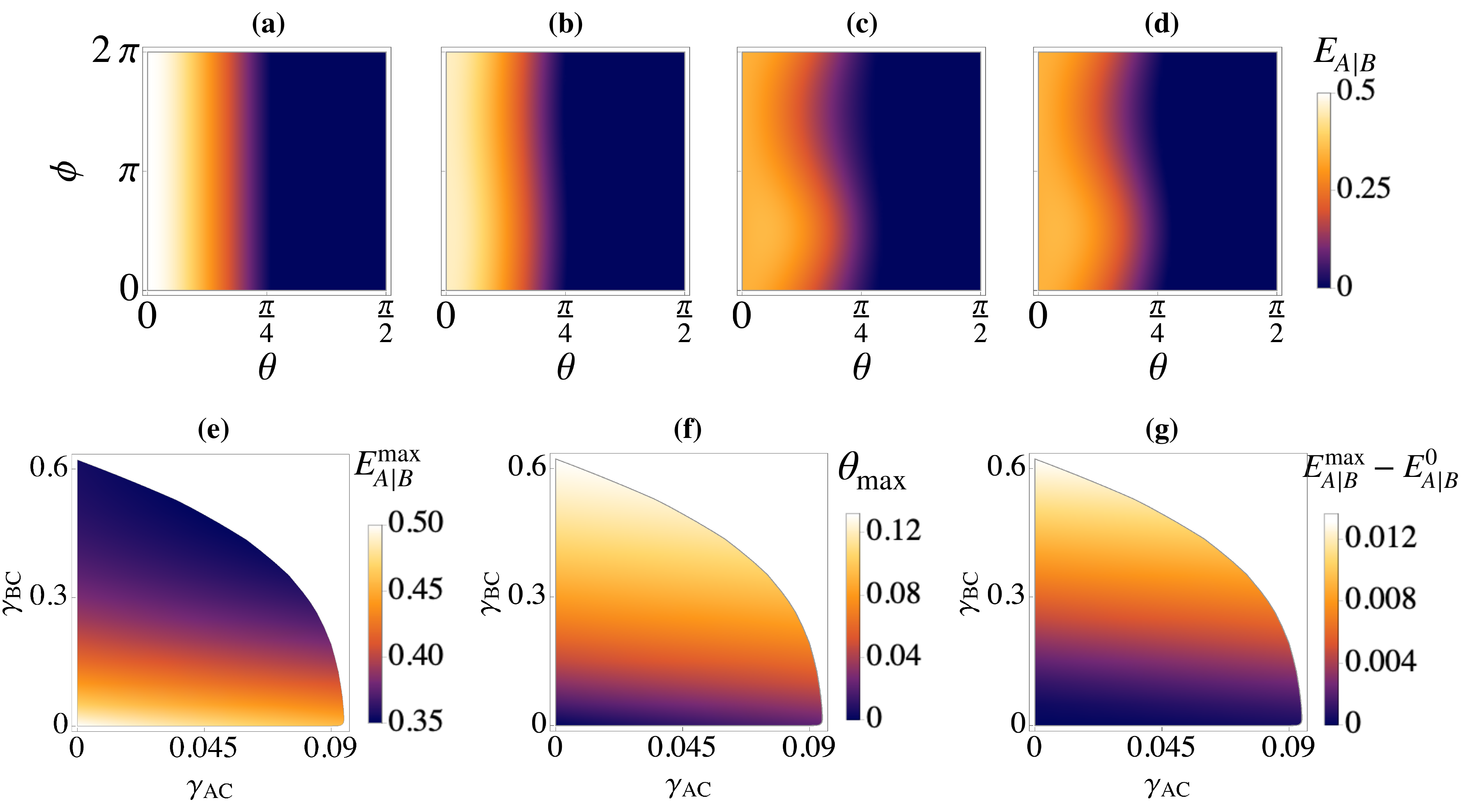}
	\caption{{\bf (a)}-{\bf (d)} Entanglement between $A$ and $B$ after measuring $C$ using a projective measurement $\Pi_C = \ket{\psi} \bra{\psi}_C$ where $\ket{\psi}_C = \cos{\theta} \ket{0}_C + e^{i \phi} \sin{\theta} \ket{1}_C$. In panel {\bf (a)}, we show the case where $\gamma_{AC}=\gamma_{BC}=0$, in {\bf (b)} $\gamma_{AC}=0.09$ and $\gamma_{BC}=0$, in {\bf (c)} $\gamma_{AC}=0$ and $\gamma_{BC}=0.6$ and in {\bf (d)} $\gamma_{AC}=0.06$ and $\gamma_{BC}=0.4$. {\bf (e)} Maximum entanglement that can be generated between $A$ and $B$ when the projective measurement $\Pi_C$ is performed on $C$. {\bf (f)} Value of $\theta$ which maximises entanglement $E_{A|B}$ when $C$ is measured with a projective measurement. In this case $E_{A|B}$ is maximised over both $\theta$ and $\phi$. {\bf (g)} Difference in entanglement generated between $A$ and $B$ when maxmised over all possible projective measurements on $C$ and when $C$ is measured in the standard basis and the state $\ket{0}_C$ is post-selected.}
	\label{fig:optimalMeasure}
\end{figure*}

However, we can take a different strategy and instead measure the state of the carrier $C$. We consider a general projective measurement described by the projector $\Pi_C=\ket{\psi} \bra{\psi}_C$ with $\ket{\psi}_C = \cos{\theta} \ket{0}_C + e^{i \phi} \sin{\theta} \ket{1}_C$ ($\theta\in[0,\pi/2]$, $\phi\in[0,2\pi]$) and investigate the effect of the parameters $\theta$ and $\phi$ on the amount of entanglement between $A$ and $B$ at the end of the protocol. Panels {\bf (a)}-{\bf (d)} in Fig.~\ref{fig:optimalMeasure} show the results for different fixed values of $\gamma_{AC}$ and $\gamma_{BC}$. 

It is clear that by measuring $C$ we can generate much higher values of entanglement than by tracing $C$ out of the system. Physically, this means that we should perform a measurement on $C$ rather than simply neglecting it while observing $AB$. In each case considered in Fig.~\ref{fig:optimalMeasure} {\bf (a)}-{\bf (d)}, it is possible to achieve more than 8 times the amount of entanglement between $A$ and $B$ than when $C$ is traced out. This is more evident in Fig.~\ref{fig:optimalMeasure} {\bf (e)} where we show the maximum entanglement generated in $A$ and $B$ through measuring $C$. This is found for the values of $\gamma_{AC}$ and $\gamma_{BC}$ which allow for EDSS as presented in Sec.~\ref{sec:encodingMoreRestricted}. The lowest value of $E_{A|B}$ is 0.348, compared to a maximum value of 0.0427 in Fig.~\ref{fig:measTraceC} where $C$ is traced out of the system.  Therefore, measuring $C$ is the better method of extracting entanglement between $A$ and $B$ for our protocol.

Additionally, Fig.~\ref{fig:optimalMeasure} {\bf (e)} shows the maximum entanglement for different strengths of incoherent dynamics in the encoding and decoding operations (we remind that the state of $AB$ is maximally entangled when $E_{A|B}=0.5$). We can notice that a large amount of entanglement can be generated even when the dynamics of the protocol features a large amount of incoherence. Therefore the protocol is robust against incoherent dynamics in both the encoding and decoding operations. 

As a result, we can meet our aim as set out in Sec.~\ref{sec:setup} of gaining entanglement between $A$ and $B$ while the carrier remains separable from $AB$ throughout the protocol. In comparing the entanglement in the final state with that in the initial state, for any $\gamma_{AC}$ and $\gamma_{BC}$ it is possible to achieve an entanglement gain of at least 0.331.

The amount of entanglement produced is heavily dependent on the value of $\theta$. In each of the plots in Fig.~\ref{fig:optimalMeasure} {\bf (a)}-{\bf (d)}, when $\theta=0$ we achieve high entanglement but when $\theta=\pi/2$, $A$ and $B$ are separable. Therefore, if we measure $C$ in the standard basis $\{ \ket{0}_C, \ket{1}_C \}$ and post-select the state $\ket{0}$ then the protocol is successful. This is unsurprising; in the specific protocol addressed here (which is based on the analysis in Refs.~\cite{cubitt} and \cite{alphaPaper}), this measurement results in a maximally entangled state of $A$ and $B$ when there are no incoherent dynamics. 

When the encoding and decoding operations take the form of CNOT operations with no imperfections, measurement in the standard basis is optimal. When the incoherent dynamics in the system are strong, however, it is unclear which measurement is optimal and results in the largest amount of entanglement. To investigate this, we identified the value of $\theta$ which produces the maximum entanglement for each value of $\gamma_{AC}$ and $\gamma_{BC}$ which allow $C$ and $AB$ to be separable throughout the protocol. The results are shown in Fig.~\ref{fig:optimalMeasure} {\bf (f)}. The plot shows that $\theta$ is always small, so the optimal measurement is close to measurement in the standard basis with post-selection of the state $\ket{0}$. Nonetheless, measurement in the computational basis is not optimal. The optimal measurement becomes further apart from the standard basis measurement as the strengths of the incoherent dynamics increase.

To compare the optimal measurement with measurement in the computational basis, we also plotted the difference in entanglement in Fig.~\ref{fig:optimalMeasure} {\bf (g)}. The difference grows as the strength of the incoherent dynamics increases, but it is small; the maximum difference is 0.0132.  Therefore measurement in the computational basis, while not optimal, is still a useful measurement in terms of extracting entanglement when state $\ket{0}$ is post-selected. It also allows us to more directly compare our results with those in Refs.~\cite{cubitt} and \cite{alphaPaper}. 

In the rest of this paper, therefore, we will measure $C$ to extract entanglement $E_{A|B}$ from the state of the system $ABC$ after decoding. We will both find the entanglement when the standard measurement is used and optimise the results over all possible projective measurements.

\section{Effect of unitary errors} \label{sec:varyUnitaryStrength}

Besides the imperfections introduced by the incoherent dynamics, the performances of the EDSS protocol can be spoiled by the coherent dynamics as well, if they fail to perfectly implement the ideal CNOT. We define these unitary errors as imperfections in gate operations which may result in a different, yet still unitary, operation being performed than the one desired.

We can study the effect of these unitary errors by varying the strength of the unitary dynamics in both encoding and decoding steps. This enables us to simulate this type of imperfections so that the unitary dynamics remain unitary but do not carry out a perfect CNOT operation. To investigate this, we rescale the strength of $H_{AC,BC}$ by a parameter $\beta_{AC,BC}$ so that the dynamical maps in Eqs.~(\ref{eq:encoding}) and \eqref{eq:decoding} become
\begin{equation} \label{eq:unitaryStrength}
	{\dot\rho} = - i [\beta_{jC} H_{jC}, \rho] + \gamma_{jC} \mathcal{L}_{jC} (\rho)
\end{equation}
with $j=A,B$. Notice that, given that we consider a finite interaction time ($t_{jC}=1$, see Sec.~\ref{sec:times}), the dynamics generated by Eq. (\ref{eq:unitaryStrength}) cannot be accounted for simply by rescaling the parameters in Eqs. (\ref{eq:encoding}) and (\ref{eq:decoding}). We first consider  the effects on the encoding and decoding steps separately, before evaluating how much entanglement we can distribute by measuring the carrier system $C$.

\subsection{Encoding: effects of coherent interaction strength}

\begin{figure}[t]
	\centering
	\includegraphics[width=.95\linewidth]{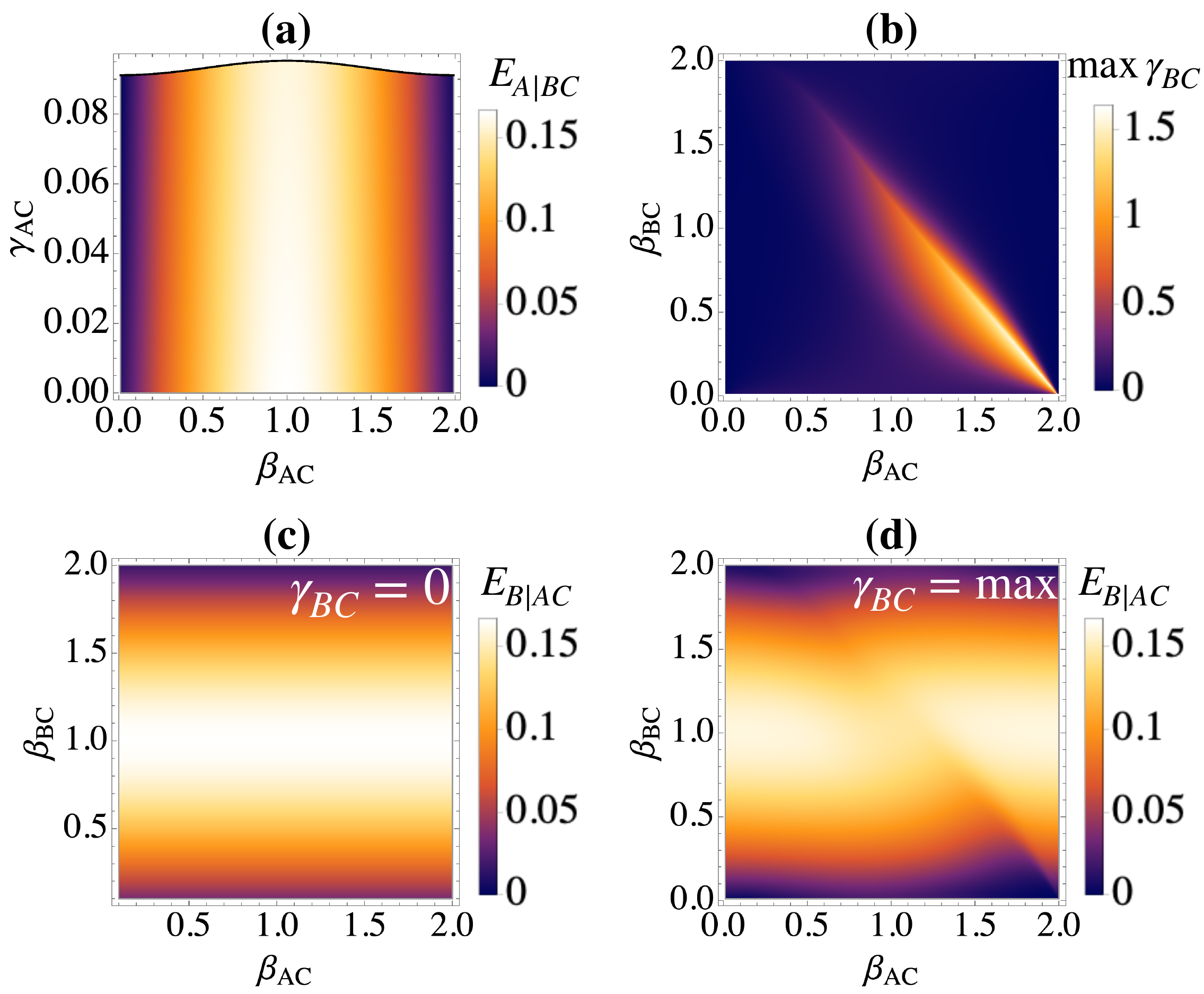}
	\caption{{\bf (a)} Entanglement between $A$ and $BC$ after the encoding step, plotted for values of $\gamma_{AC}$ which do not cause $C$ to become entangled with $AB$. The maximum value that $\gamma_{AC}$ can take is shown by the solid black line at the top of the plot. Entanglement is highest when $\beta_{AC}=1$. {\bf (b)} Maximum value of $\gamma_{BC}$ which allows the carrier to remain separable from the rest of the system (taking $\gamma_{AC}=0$). {\bf (c)}-{\bf (d)} Entanglement between $B$ and $AC$ after the decoding step when {\bf (c)} $\gamma_{AC}=\gamma_{BC}=0$, {\bf (d)} $\gamma_{AC}=0, \gamma_{BC}$ is the maximum value it can take so that $C$ and $AB$ remain separable throughout the decoding step.}
	\label{fig:betaEncDec}
\end{figure}

The effect of varying the strength of the unitary dynamics in the encoding operation is shown in Fig.~\ref{fig:betaEncDec} {\bf (a)}. We show the values of $\beta_{AC}$ and $\gamma_{AC}$ for which $C$ and $AB$ are always separable during the encoding step. The impact that $\beta_{AC}$ has on the values we can use for $\gamma_{AC}$ is very small in this case; no matter what the value of $\beta_{AC}$, $\gamma_{AC}$ can always take a maximum value between 0.09 and 0.095. 

The density plot itself in Fig.~\ref{fig:betaEncDec} {\bf (a)} shows the entanglement between $A$ and $BC$ after the encoding step. This effectively measures the  ``success" of the encoding step. Here, the value of $\beta_{AC}$ has a significant effect. As can be clearly seen in Fig.~\ref{fig:betaEncDec} {\bf (a)}, entanglement is highest when $\beta_{AC}\simeq1$ and, indeed, the maximum entanglement is obtained when $\beta_{AC}=1$. This is unsurprising: for $\beta_{AC}=1$, given that the evolution time of the encoding step is $t_{AC}=1$, the unitary part of the dynamics in Eq.~(\ref{eq:unitaryStrength}) effectively allows a perfect CNOT operation to be carried out between $A$ and $C$. Therefore, when this value is changed (or we change the duration of the interaction), the CNOT operation is only imperfectly completed and less entanglement is generated between $A$ and $B$. 

The value of $\gamma_{AC}$ also has an effect on $A|BC$ entanglement; as $\gamma_{AC}$ increases, $E_{A|BC}$ decreases. However, this effect is small and a large amount of entanglement can be generated for any value of $\gamma_{AC}$ as long as $\beta_{AC}\simeq1$.

\subsection{Decoding: effects of coherent interaction strength}

In the case of decoding, we first analyze how the strength of the unitary dynamics in the encoding and decoding steps affects the strength of the incoherent dynamics. The maximum value that can be taken for $\gamma_{BC}$ is plotted in Fig.~\ref{fig:betaEncDec} {\bf (b)} against the unitary strengths when $\gamma_{AC}=0$. In this plot we can see that varying the unitary strength can greatly increase or decrease the range of values of $\gamma_{BC}$ for which the carrier is separable throughout the protocol. Whenever the strength of the unitary dynamics in the encoding operation increases and the decoding strength decreases, then $\gamma_{BC}$ can be allowed to assume values larger than 1 before $C$ becomes entangled with $AB$. Therefore, if we have very strong incoherent dynamics in our system, we can still achieve EDSS if we can modify the strengths of the unitary dynamics in the protocol. 

In Figs.~\ref{fig:betaEncDec} {\bf (c)} and \ref{fig:betaEncDec} {\bf (d)}, we study how well the decoding operation works, i.e. how much entanglement we can produce between $B$ and $AC$. Fig.~\ref{fig:betaEncDec} {\bf (c)} shows $E_{B|AC}$ whenever $\gamma_{BC}=0$ and it is clear that the closer $\beta_{BC}$ is to 1, the higher the amount of entanglement generated between $B$ and $AC$. This is similar to the case of encoding dynamics in Fig.~\ref{fig:betaEncDec} {\bf (a)}.

Fig.~\ref{fig:betaEncDec} {\bf (d)} shows $E_{B|AC}$ whenever $\gamma_{BC}$ takes its maximum value as plotted in Fig.~\ref{fig:betaEncDec} {\bf (b)}. We notice that it is still possible to generate large amounts of entanglement whenever the incoherent dynamics are strong. Additionally, we see that in the cases where $\gamma_{BC}$ can take particularly large values, the entanglement decreases but it is not destroyed completely.

In both plots, we find that the maxima of Figs.~\ref{fig:betaEncDec} {\bf (c)} and \ref{fig:betaEncDec} {\bf (d)} do not overlap with the maximum of Fig.~\ref{fig:betaEncDec} {\bf (b)}; there is a compromise to be made between how much entanglement can be generated and how flexible we can be in the strength of the incoherent dynamics. Therefore depending on what we need more, i.e. robustness against incoherent dynamics or as much entanglement distributed as possible, we can tune the strengths of the unitary dynamics to achieve our aims.

\subsection{Maximum entanglement when the carrier is measured}

\begin{figure}[t!]
	\includegraphics[width=1\columnwidth]{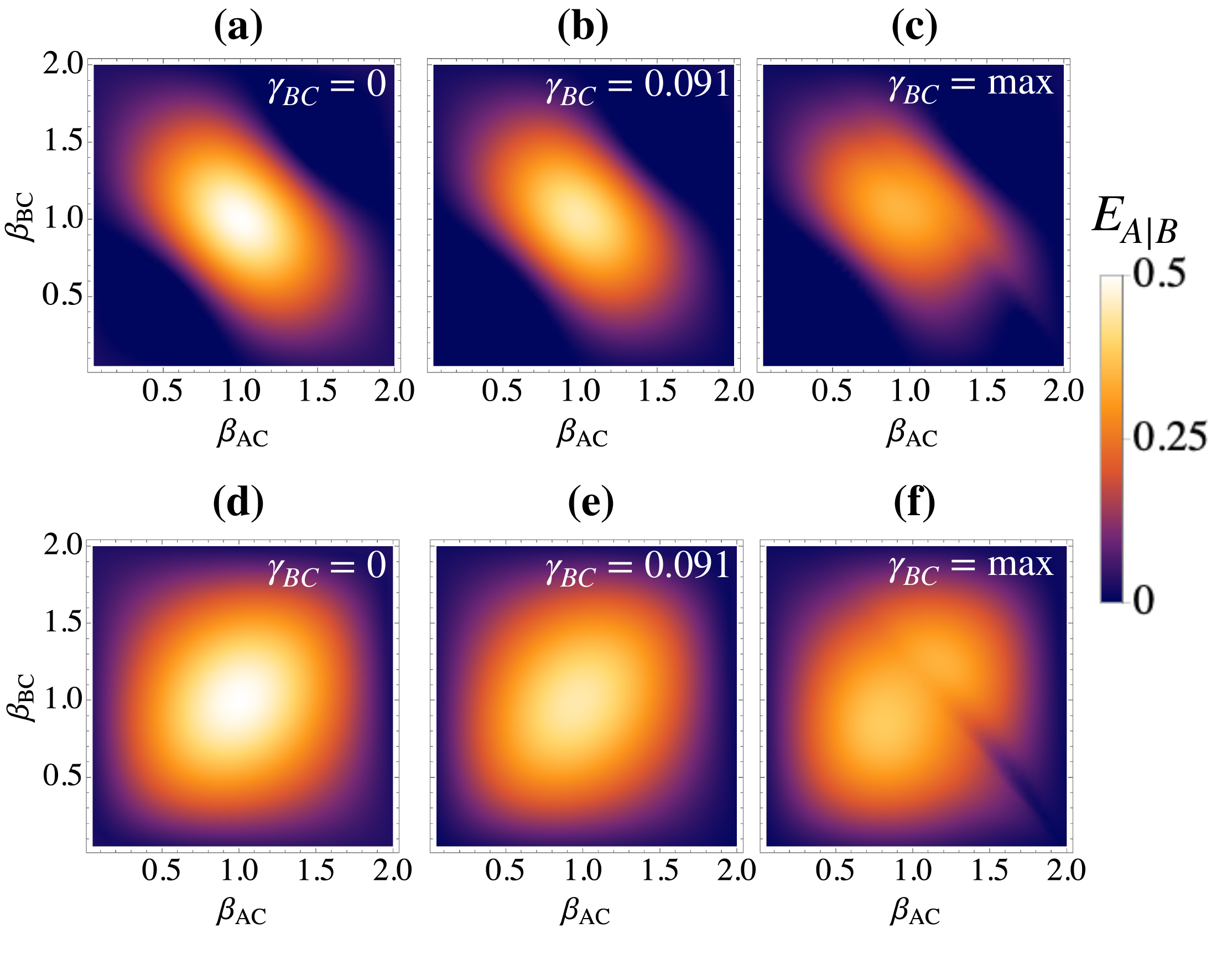}
	\caption{{\bf (a)}-{\bf (c)} Entanglement between $A$ and $B$ after measuring $C$ in the standard basis and post-selecting the state $\ket{0}$. We take $\gamma_{AC}=0$ and {\bf (a)} $\gamma_{BC}=0$, {\bf (b)} $\gamma_{BC}=0.091$, the highest value for which $C$ and $AB$ are separable for every value of $\beta_{AC}$ and $\beta_{BC}$ and {\bf (c)} the maximum value of $\gamma_{BC}$ for which $C$ and $AB$ are separable for each value of $\beta_{AC}$ and $\beta_{BC}$. {\bf (d)}-{\bf (f)} Maximum entanglement between $A$ and $B$ after measuring $C$ with the projector $\Pi_C$. The entanglement is maximised over $\theta$ and $\phi$ and we take $\gamma_{AC}=0$ and {\bf (d)} $\gamma_{BC}=0$, {\bf (e)} $\gamma_{BC}=0.091$ and {\bf (f)} the maximum value of $\gamma_{BC}$ for which $C$ and $AB$ are separable for each value of $\beta_{AC}$ and $\beta_{BC}$.}
	\label{fig:varyBetaMeasure}
\end{figure}

So far we have gained an idea of which values of $\beta_{AC}$ and $\beta_{BC}$ will give the highest entanglement; from Figs.~\ref{fig:betaEncDec} {\bf (a)} and \ref{fig:betaEncDec} {\bf (c)}, we can see that both $A|BC$ entanglement after encoding and $B|AC$ entanglement after decoding are highest whenever $\beta_{AC}=\beta_{BC}=1.$ It is now left to investigate the effect of $\beta_{AC}$ and $\beta_{BC}$ on $A|B$ entanglement whenever $C$ is measured. 

In Figs.~\ref{fig:varyBetaMeasure} {\bf (a)}-{\bf (c)}, we can see how the entanglement, after measuring $C$ in the standard basis, is affected by increasing the strength of the incoherent dynamics $\gamma_{BC}$. The state $\ket{0}$ is post-selected as before (see Sec.~\ref{sec:doesntPreventEDSS}). In Fig.~\ref{fig:varyBetaMeasure} {\bf (a)}, where $\gamma_{BC}=0$, we can achieve a maximally entangled state of $AB$ when $\beta_{AC}=\beta_{BC}=1$. In Fig.~\ref{fig:varyBetaMeasure} {\bf (b)}, we took $\gamma_{BC}=0.091$; this is the largest value that we can use for $\gamma_{BC}$ so that for all values of $\beta_{AC}$ and $\beta_{BC}$ in the interval $0 \leq \beta_{AC}, \beta_{BC} \leq 2$, $C$ and $AB$ remain separable. The entanglement between $A$ and $B$ here is lower than in Fig.~\ref{fig:varyBetaMeasure} {\bf (a)}, but is still significant. For instance, the maximum entanglement (when $\beta_{AC}=\beta_{BC}=1$) is 0.44. Finally, Fig.~\ref{fig:varyBetaMeasure} {\bf (c)} shows the entanglement when, for each value of $\beta_{AC}$ and $\beta_{BC}$, $\gamma_{BC}$ takes its maximum value as shown in Fig.~\ref{fig:betaEncDec} {\bf (b)}. Again, the entanglement decreases when compared to the first two plots. However, it does not vanish completely. Indeed, the maximum entanglement, occurring when $\beta_{AC}=\beta_{BC}=1$, is 0.34. {This implies a significant maximum entanglement gain of 0.323 since the start of the protocol.}

One interesting feature of the three plots is how they are aligned along the diagonal $\beta_{AC}+\beta_{BC}=2$. Clearly the highest entanglement is achieved when $\beta_{AC}=\beta_{BC}=1$, but if this is not possible then it is better if $\beta_{AC}$ increases and $\beta_{BC}$ decreases, or vice versa. This is due to the fact that when $\beta_{AC}+\beta_{BC}=2$, there is an element in the state $ABC$ of the form $\ket{\phi^+}_{AB} \bra{\phi^+} \otimes \ket{0}_{C} \bra{0}$. When $\beta_{AC} = \beta_{BC} = 1$, this is the only element which remains when $C$ is measured in the standard basis and outcome 0 is obtained. As $\beta_{AC}$ and $\beta_{BC}$ get further from 1, then more elements are added which, when $C$ is measured and outcome 0 is obtained, are included in the final state of $AB$ and so the entanglement between $A$ and $B$ decreases. 

Figs.~\ref{fig:varyBetaMeasure} {\bf (d)}-{\bf (f)} present entanglement between $A$ and $B$ maximised over all possible projective measurements on $C$. We see that high entanglement is no longer limited to values of $\beta_{AC}$ and $\beta_{BC}$ which are close to the line $\beta_{AC}+\beta_{BC}=2$. Instead, the amount of entanglement which can be generated in this way depends on how close $\beta_{AC}$ and $\beta_{BC}$ are to 1. If $\beta_{AC}$ or $\beta_{BC}$ is close to 0 or 2, then the unitary part of the dynamics effectively vanishes. In comparing Figs.~\ref{fig:varyBetaMeasure} {\bf (a)} and {\bf (d)}, especially the bottom left and top right hand corners, we see that the standard basis measurement can be very far from optimal in terms of entanglement as the strengths of the unitary dynamics change. 

In Fig.~\ref{fig:varyBetaMeasure} {\bf (f)}, where for each pair of values $(\beta_{AC}, \beta_{BC})$, the strength of the incoherent dynamics in the decoding operation $\gamma_{BC}$ takes the maximum value which allows for EDSS, we see that the maximum entanglement is no longer centred on $(1,1)$. The entanglement is highest ($E_{A|B} = 0.384$) when $\beta_{AC} = 0.8$ and $\beta_{BC} = 0.85$ since the maximum strength of incoherent dynamics is relatively small ($\gamma_{BC} = 0.282$). 

There is a small region in the bottom right hand corner of Fig.~\ref{fig:varyBetaMeasure} {\bf (f)} where $\gamma_{BC}$ takes very high values; in this case it is not possible to generate large entanglement. We would need to reduce the strength of the incoherent dynamics in order to produce higher entanglement for these values of $\beta_{AC}$ and $\beta_{BC}$. Nevertheless, aside from this region, the restrictions placed on $\gamma_{BC}$ by the separability condition ensure large entanglement gain in many cases. 

Indeed, let us consider our original goal of using a separable carrier $C$ to increase entanglement between $A$ and $B$. Since the initial entanglement between $A$ and $B$ is only 0.0167, Fig.~\ref{fig:varyBetaMeasure} {\bf (f)} shows that we would require substantial unitary errors to lose $AB$ entanglement during the protocol even in the presence of strong incoherent dynamics. For instance, the entanglement in the region given by $\beta_{AC},\beta_{BC}=1\pm0.7$ is always higher than the initial $AB$ entanglement. Consequently, the encoding and decoding steps may be significantly different from unitary CNOT operations and yet still allow us to meet our goal and distribute entanglement between the remote subsystems.

\section{Conclusions} \label{sec:conc}

We have shown that entanglement distribution with separable carriers is still possible when we cannot perform unitary operations perfectly, due to either non-unitary or unitary errors. In particular, we have focused on a version of EDSS based on CNOT encoding and decoding steps. In the ideal implementation of such protocol, these choices would correspond to a maximally entangled state of the remote subsystems $A$ and $B$. Even though, under the effects of encoding and decoding operations, maximally entangled states can no longer be generated, the amount of entanglement between $A$ and $B$ can still be very large when $C$ is measured. This is a crucial feature of EDSS that would take the protocol closer to experimental validation in settings where unitary operations cannot be carried out without imperfections. In addition, this opens avenues for practical quantum information implementations such as photonic schemes with semiconductor sources of quantum light for quantum internet or quantum computation schemes (in an integrated environment maybe including heralded photonic states). High fidelities are indeed possible in this type of scheme, but they are also unavoidably affected by experimental imperfections and the source's lack of full ideality \cite{chung16,bassobasset19,llewellyn20}. In general, whether EDSS would be more advantageous than a direct entanglement-generation protocol is a strongly platform-dependent issue that needs to be addressed on a case-by-case basis. The scenario depicted above, though, appears to be a significant instance of a physical system that would benefit from indirect approaches such as EDSS.

An interesting conceptual extension for these imperfect photonic states would be the possibility of exploiting EDSS in quantum gate operations to maintain a non-ideal but somehow constant level of entanglement through a tailored photonic circuit; the (heralded) EDSS scheme could effectively counter the depletion of entanglement between two remote information carriers through multiple interactions with a ``correcting" third party. 

\begin{acknowledgements}
The  authors  acknowledge  financial  support from H2020 through the Collaborative Project TEQ (Grant Agreement  No.766900), the DfE-SFI Investigator Programme (Grant No. 15/IA/2864),  the  Leverhulme  Trust  Research  Project Grant  UltraQute  (grant  nr.  RGP-2018-266),  COST  Action CA15220, and the Royal Society Wolfson Research Fellowship scheme (RSWF$\backslash$R3$\backslash$183013).
\end{acknowledgements}

\section*{Conflict of interest}
The authors declare that they have no conflict of interest.

%
%

\end{document}